# An Analytical Formula for Determining the Electrical Impedance between a Single Adherent Cell and Sensor Substrate

Masataka Shiozawa[1]* and Shigeyasu Uno[1]*

[1]*Department of Electrical Systems, Graduate School of Science of Engineering, Ritsumeikan University, Shiga 525-8577, Japan*

E-mail: re0071sr@ed.ritsumei.ac.jp; suno@fc.ritsumei.ac.jp

## Abstract

An analytical formula for electrical impedance between an adherent living cell and sensor substrate measured using a microelectrode is presented for the first time. Previously-reported formula has been applicable only for the case where many cells are on a large electrode. In contrast, our formula is valid even when a microelectrode smaller than the cell size is underneath the cell, which is often the case for the state-of-the-art single-cell analysis. Numerical simulations for verifying the accuracy of our formula reveals that the discrepancies between the theoretical impedances calculated by our formula and numerical simulation results are negligibly small. Our formula will be useful for describing cell-substrate impedance properties in equivalent circuit model analysis or sensor design optimizations.

# 1. Introduction

In recent years, a number of investigations have revealed that an understanding of living cell behaviors is particularly important for new drug development and early cancer detection[1-4]. Therefore, living cell analyses have applications to drug discovery and disease diagnosis. The numerous methods for living cell analyses have been proposed in previous works. Raman spectroscopy, quantitative phase microscopy and flow cytometry are some of optical measurements[5–7]. However, these optical methods require the cells to be labeled for staining and the measurement systems are large. On the other hand, quartz crystal microbalance and surface acoustic wave sensors do not require cell labeling[8,9]. These methods are non-invasive, and the measurement systems can be much smaller than the optical systems. However, they are not suitable to detect the changes in cell morphology or cell-to-cell junction quality. Electrochemical measurements have long been used to quantitatively analyze morphological changes of cells; these measurements have some advantages such as non-invasion, real time, small measurement system, and no-labeling[10–13]. There are three major electrochemical measurement methods. First, the potentiometry has the capability of measuring potential changes caused by oxidation-reduction reaction. Glucose sensor and pH sensor, which are used in cell analysis, employ this measurement method[14–16]. In addition, this measurement method enables to measure changes in the action potential which neuron and muscle cells cause[17–19]. Second, the amperometry is the way of measuring the electro-activity of oxidation-reduction reaction on the interface as electrical current. Some studies reported that the amperometry sensor can detect molecules released from cells[20–22]. Third, the impedance measurement is a non-invasive way of detecting the changes in ionic current paths around cells, conductivity of cytoplasm, and cell membrane permittivity[11,13]. Previous studies showed that the impedance measurement has the capability to detect morphological change, evaluation of toxicity, cell junction, cell adhesion, cell proliferation, apoptosis, and necrosis[11,13,23–25]. In particular, electrochemical impedance spectroscopy (EIS), which is the impedance measurement by sweeping frequency, provides various information such as morphological and electrical cell changes simultaneously[13,25]. Hence EIS is optimal for living cell analyses. In previous works, various electrode structures were proposed for electrochemical living cell analyses in the impedance measurement. Planar interdigitated electrode arrays are fabricated at low cost, and the theory and simulation methods are well-established[26,27]. Parallel facing electrodes are used in the flow cytometry for analyzing single cells[28,29], and this electrode structure has the capability of screening single cells with high throughput[29]. Microelectrode arrays can identify where a target cell exists and map it at moderate resolution[30,31]. In addition, the use of microelectrode arrays enables the whole measurement system downsizing through the application of complementary metal-oxide-semiconductor (CMOS) technology[32–35]. As with parallel facing electrodes, microelectrodes achieve high throughput screening by using CMOS technology. Besides, their sensitivity is in general higher than that of conventional electrode structures. Thus, microelectrodes are attracting considerable attention. As mentioned above, EIS measurement using microelectrodes has been expected to play important roles in living cell analyses. In EIS measurement, equivalent circuit model (ECM) is employed for quantitative analyses of various cell changes from the impedance change[36–38]. An appropriate analytical formula for ECM is essential to obtain the parameters such as conductivity, permittivity, and cell size by fitting experimental results. The analytical formula also provides the insights in the most optimized design of the sensor geometry. In general, cell-to-cell interactions should affect original characteristics of cell populations and tissues. Thus, single-cell analysis taking into account of the heterogeneous behaviors of a single cell will be appropriate[39–41]. There are some studies using microelectrodes with their size ranging from 1.0 μm to 8.0 μm used for cell impedance mapping[33,42] and some others reported theoretical analysis of single-cell behavior when using such electrodes with computer simulation[42,43]. However, there is no report on the appropriate analytical formula for cell-substrate impedance with such small electrodes. As noted above, in general, appropriate formulae are essential for quantitative analyses and the several parameters which describe properly cell characteristics in living cell analyses. Giaever et al. proposed a formula for describing the impedance between an adherent cell and electrode[44]. Nevertheless, it was assumed that the electrode was large; this formula is inapplicable when the electrode is smaller than the cell.

In this work, we derived an analytical formula for cell-substrate impedance when the electrode is smaller than the cell. The accuracy of the formula was verified through two numerical simulations. We swept the electrode radius in the first simulation and the height between cell and substrate in the second simulation. From the first simulation, we observed negligible error between the theoretical value from analytic formula and numerical simulation result. The second simulation showed that our formula well-described cell-substrate impedance when the cell adheres to the substrate. Thus, our formula can describe cell-substrate impedance when the electrode is smaller than the cell. It will be useful for the experimentalists in identifying cell behaviors and designing optimized sensor structures.

## 2. Theory

### 2.1. Giaever's Theory

Giaever et al. proposed a formula for describing the impedance between an adherent cell and electrode as shown in Fig. 1, where $z$ and $r$ are the height and radius, respectively. Note that this formula was derived under the assumption that the electrode was sufficiently larger than the cell. In Fig. 1, the electric potential $V(r)$ and the current $I(r)$ as functions of the radius $r$ are expressed by the following equations[44],

$$V(r) = -\frac{1}{Z_m + Z_n} \frac{Z_m V_e + Z_n V_c}{I_0(\gamma_A r_c) + 2h\sigma_{sol}\gamma_A \frac{R_b}{r_c} I_1(\gamma_A r_c)} I_0(\gamma_A r) + \frac{1}{Z_m + Z_n}(Z_m V_e + Z_n V_c), (1)$$

$$I(r) = 2\pi r h \sigma_{sol} \frac{Z_m V_e + Z_n V_c}{Z_m + Z_n} \frac{\gamma_A}{I_0(\gamma_A r_c) + 2h\sigma_{sol}\gamma_A \frac{R_b}{r_c} I_1(\gamma_A r_c)} I_1(\gamma_A r), \quad (2)$$

where $V_e$ [V] and $V_c$ [V] indicate the electric potentials at the electrode and the cytoplasm, respectively. $I_n(x)$ is a modified Bessel function of the first kind of order $n$, with non-dimensional variable $x$. $\sigma_{sol}$ [Sm$^{-1}$] indicates the conductivity of the solution, and $r_c$ [m] and $h$ [m] are the radius of the cell and the distance between the cell and substrate, respectively. $R_b$ [$\Omega$m$^2$] is specific resistance between cells per unit area when several cells exist on the electrodes, and $Z_n$ [$\Omega$m$^2$] is the specific impedance of the electrode, as described by the following equation,

$$Z_n = \frac{1}{j\omega C_{dl0}}, \quad (3)$$

where $j$ and $\omega$ [s$^{-1}$] are the imaginary unit and angular frequency, respectively, and $C_{dl0}$ [Fm$^{-2}$] is the electrical double-layer capacitance per unit area, given by

$$C_{dl0} = \frac{\epsilon_0 \epsilon_{sol}}{\lambda_D}. \quad (4)$$

Here, $\epsilon_0$ [Fm$^{-1}$], $\epsilon_{sol}$, and $\lambda_D$ [m] are the permittivity of the vacuum, relative permittivity of the medium solution, and the Debye length, respectively. $Z_m$ [$\Omega$m$^2$] is the specific impedance of the cell membrane, as expressed by the following equation,

$$Z_m = \frac{1}{j\omega C_{mem0}}, \quad (5)$$

where $C_{mem0}$ [Fm$^{-2}$] is the capacitance of the cell membrane per unit area, given by

$$C_{mem0} = \frac{\epsilon_0 \epsilon_{mem}}{d_{mem}}. \quad (6)$$

Here, $\epsilon_{mem}$ and $d_{mem}$ [m] are the relative permittivity and thickness of the cell membrane, respectively. $\gamma_A$ [m$^{-1}$] is defined as

$$\gamma_A = \sqrt{\frac{1}{h\sigma_{sol}} \left( \frac{1}{Z_n} + \frac{1}{Z_m} \right)}. \quad (7)$$

Assuming that electric potential drop occurs mainly under the cell, $V_c$ can be set to zero. Thus, the following equation represents $Z_{Giaever}$ [$\Omega$m$^2$] using the total currents flowing out of the working electrode (WE)

$$Z_{\text{Giaever}}^{-1} = \frac{1}{Z_n}\left(\frac{Z_n}{Z_m + Z_n} + \frac{Z_m}{Z_m + Z_n}\frac{1}{\frac{r_c\gamma_A}{2}\frac{I_0(\gamma_A r_c)}{I_1(\gamma_A r_c)} + R_b\left(\frac{1}{Z_n} + \frac{1}{Z_m}\right)}\right). \tag{8}$$

If only a single cell is on the electrode, $R_b$ becomes zero because it is associated with intercellular resistance between many cells[44].

2.2. Formula describing cell-substrate impedance when using a microelectrode

Fig. 2 shows schematics of a situation where an electrode smaller than the cell is used for single-cell analysis. In Fig. 2(a), cell-substrate area is divided into two regions A and B, and the formula is derived from the relationship between the potential and current over an infinitesimal distance in each region. Figs. 2(b) and (c) represent the relationships between potentials and currents in regions A and B in Fig. 2(a), respectively. For simplicity, we assume that the potential in cell-substrate gap depends only on $r$, and neglect the potential drop along the $z$-axis. This is reasonable because the distance between the cell and substrate is sufficiently small. In region A in Fig. 2(a), the relationship between the potential and the current is expressed by Giaever's model as follows[44],

$$-dV_A = I_A \frac{1}{2\pi r h \sigma_{\text{sol}}} dr, \tag{9}$$

$$V_e - V_A = \frac{Z_n}{2\pi r dr} dI_e, \tag{10}$$

$$V_A - V_c = \frac{Z_m}{2\pi r dr} dI_c, \tag{11}$$

$$dI_A = dI_e - dI_c, \tag{12}$$

where $V_A(r)$ [V] and $I_A(r)$ [A] are the potential and current at the radial position $r$ in region A, respectively. In addition, $I_e(r)$ [A] and $I_c(r)$ [A] are the total currents flowing out of the WE and through the cell membrane, respectively. From Eqs. (9)–(12), we obtain the differential equation for $V_A$,

$$\frac{d^2 V_A}{dr^2} + \frac{1}{r}\frac{dV_A}{dr} - \frac{1}{h\sigma_{sol}}\left(\frac{1}{Z_n} + \frac{1}{Z_m}\right)V_A + \frac{1}{h\sigma_{sol}}\left(\frac{V_e}{Z_n} + \frac{V_c}{Z_m}\right) = 0. \tag{13}$$

The general solution to Eq. (13) is given by

$$V_A(r) = C_A I_0(\gamma_A r) + D_A K_0(\gamma_A r) + \frac{1}{Z_n + Z_m}(Z_m V_e + Z_n V_c), \tag{14}$$

where $K_n(x)$ is the modified Bessel function of the second kind of order $n$, with a non-dimensional variable $x$. $C_A$ and $D_A$ are the constants of integration determined from the boundary condition. As $K_0(\gamma_1 r)$ tends to infinity as $r$ tends to zero, $D_A$ must be set to zero, giving

$$V_A(r) = C_A I_0(\gamma_1 r) + \frac{1}{Z_n + Z_m}(Z_m V_e + Z_n V_c). \tag{15}$$

In region B in Fig. 2(a), the bottom of this region is not the WE, but rather the non-conducting substrate, i.e., $dI_e = 0$. Therefore, the differential equation and its solution for the potential in region B can be obtained by setting the $Z_n$ to infinity in the Eq. (14). The potential then becomes

$$V_B(r) = C_B I_0(\gamma_B r) + D_B K_0(\gamma_B r) + V_c, \tag{16}$$

where $V_B(r)$ [V] indicates the potential at the point of the $r$ coordinate in region B. Here, $C_B$ and $D_B$ are the constants of integration, and the $\gamma_B$ [m$^{-1}$] is defined as

$$\gamma_B = \sqrt{\frac{1}{h\sigma_{sol} Z_m}}. \tag{17}$$

We set three boundary conditions to determine the coefficients which are $C_A$, $C_B$, and $D_B$. First, we defined $r_e$ [m] and $r_c$ [m] as the WE radius and cell radius, respectively. The potential and current must be continuous at the boundary between the regions A and B ($r = r_e$) in Fig. 2(a), i.e., $V_A(r_e) = V_B(r_e)$ and $I_A(r_e) = I_B(r_e)$. Hence, we obtain the following equations

$$C_A I_0(\gamma_A r_e) - C_B I_0(\gamma_B r_e) - D_B K_0(\gamma_B r_e) + \frac{Z_m(V_e - V_c)}{Z_n + Z_m} = 0, \tag{18}$$

$$C_A \gamma_A I_1(\gamma_A r_e) - C_B \gamma_B I_1(\gamma_B r_e) + D_B \gamma_B K_1(\gamma_B r_e) = 0. \tag{19}$$

As the potential drops rapidly around the WE, the potential at the edge of the cell ($r = r_c$) is zero, i.e., $V_B(r_c) = 0$. Therefore, we obtain the following equation

$$C_B I_0(\gamma_B r_c) + D_B K_0(\gamma_B r_c) + V_c = 0. \tag{20}$$

We now introduce a coefficient matrix to solve simultaneous equations Eqs. (18)–(20) for $C_A$, $C_B$, and $D_B$,

$$\boldsymbol{A} \begin{bmatrix} C_A \\ C_B \\ D_B \end{bmatrix} = \begin{bmatrix} (1-k^2)(V_c - V_e) \\ 0 \\ -V_c \end{bmatrix}, \tag{21}$$

where $\boldsymbol{A}$ is the following coefficient matrix

$$\boldsymbol{A} = \begin{bmatrix} I_0(\gamma_A r_e) & -I_0(\gamma_B r_e) & -K_0(\gamma_B r_e) \\ I_1(\gamma_A r_e) & -k I_1(\gamma_B r_e) & k K_1(\gamma_B r_e) \\ 0 & I_0(\gamma_B r_c) & K_0(\gamma_B r_c) \end{bmatrix}, \tag{22}$$

and $k$ is defined as

$$k = \frac{\gamma_B}{\gamma_A} = \sqrt{\frac{Z_n}{Z_n + Z_m}} = \sqrt{\frac{C_{\text{mem0}}}{C_{\text{dl0}} + C_{\text{mem0}}}}. \tag{23}$$

Note that $k$ does not depend on frequency, and is constant within the range of $0 < k < 1$. The inverse of $\boldsymbol{A}$ is given by

$$\boldsymbol{A}^{-1} = \frac{1}{\Delta} \begin{bmatrix} A_{11} & A_{12} & A_{13} \\ A_{21} & A_{22} & A_{23} \\ A_{31} & A_{32} & A_{33} \end{bmatrix}, \tag{24}$$

$$\Delta = I_1(\gamma_A r_e)[I_0(\gamma_B r_e)K_0(\gamma_B r_c) - I_0(\gamma_B r_c)K_0(\gamma_B r_e)]$$
$$- k I_0(\gamma_A r_e)[I_1(\gamma_B r_e)K_0(\gamma_B r_c) + I_0(\gamma_B r_c)K_1(\gamma_B r_e)], \tag{25}$$

$$A_{11} = -k I_1(\gamma_B r_e)K_0(\gamma_B r_c) - k I_0(\gamma_B r_c)K_1(\gamma_B r_e), \tag{26}$$

$$A_{12} = -I_0(\gamma_B r_c)K_0(\gamma_B r_e) + I_0(\gamma_B r_e)K_0(\gamma_B r_c), \tag{27}$$

$$A_{13} = -k I_0(\gamma_B r_e)K_1(\gamma_B r_e) - k I_1(\gamma_B r_e)K_0(\gamma_B r_e), \tag{28}$$

$$A_{21} = -I_1(\gamma_A r_e)K_0(\gamma_B r_c), \tag{29}$$

$$A_{22} = I_0(\gamma_A r_e)K_0(\gamma_B r_c), \tag{30}$$

$$A_{23} = -I_1(\gamma_A r_e)K_0(\gamma_B r_e) - k I_0(\gamma_A r_e)K_1(\gamma_B r_e), \tag{31}$$

$$A_{31} = I_1(\gamma_A r_e)I_0(\gamma_B r_c), \tag{32}$$

$$A_{32} = -I_0(\gamma_A r_e)I_0(\gamma_B r_c), \tag{33}$$

$$A_{33} = I_0(\gamma_B r_e)I_1(\gamma_A r_e) - k I_0(\gamma_A r_e)I_1(\gamma_B r_e). \tag{34}$$

Consequently, from Eq. (21), $C_A$, $C_B$, and $D_B$ are given by

$$C_A = \frac{1}{\Delta}[(1-k^2)(V_c - V_e)A_{11} - V_c A_{13}], \tag{35}$$

$$C_B = \frac{1}{\Delta}[(1-k^2)(V_c - V_e)A_{21} - V_c A_{23}], \tag{36}$$

$$D_B = \frac{1}{\Delta}[(1-k^2)(V_c - V_e)A_{31} - V_c A_{33}]. \tag{37}$$

Based on these results, we derive the formula for cell-substrate impedance. The potential at the edge of the cell is set to zero as a boundary condition of Eq. (20), and cell-substrate impedance is calculated using

$$Z_c = \frac{V_e}{I_e}, \tag{38}$$

where $Z_c$ [Ω] is the impedance between the cell and substrate. $I_e$ is derived via integration of $dI_e$ in Eq. (10) with respect to $r$ as follows

$$I_e = \int_0^{r_e} dI_e = \frac{2\pi}{Z_n} \int_0^{r_e} r[V_e - V_A(r)]dr. \tag{39}$$

Finally, we derive the cell-substrate impedance as

$$Z_c^{-1} = -\frac{2\pi r_e C_A I_1(\gamma_A r_e)}{Z_n \gamma_A V_e} + \pi r_e^2 \frac{V_e - V_c}{V_e(Z_n + Z_m)}, \tag{40}$$

where $C_A$ is given by Eq. (35).

2.3. Comparison with Giaever's formula

The derived formula Eq. (40) must be the same as Giaever's formula Eq. (8) when $r_e = r_c$. With $V_c = 0$, the limit of $C_A$ as $r_e \to r_c$ is given by

$$\lim_{r_e \to r_c} C_A = -\frac{1}{I_0(\gamma_A r_c)} V_e(1 - k^2)$$

$$= -\frac{1}{I_0(\gamma_A r_c)} \frac{Z_m}{Z_n + Z_m} V_e. \tag{41}$$

Consequently, the formula of cell-substrate impedance for a unit area is described by the following equation

$$\lim_{r_e \to r_c} \frac{1}{\pi r_c^2} Z_c^{-1} = \frac{1}{\pi r_c^2} \left( \frac{2\pi r_c}{Z_n \gamma_A} \frac{I_1(\gamma_A r_c)}{I_0(\gamma_A r_c)} \frac{Z_m}{Z_n + Z_m} + \pi r_c^2 \frac{1}{Z_n + Z_m} \right)$$
$$= \frac{1}{Z_n} \left( \frac{Z_n}{Z_n + Z_m} + \frac{Z_m}{Z_n + Z_m} \frac{2 I_1(\gamma_A r_c)}{\gamma_A r_c I_0(\gamma_A r_c)} \right) = Z_{\text{Giaever}}^{-1}. \tag{42}$$

Thus, $Z_c$ for a unit area equals that given by Giaever's formula in the limit $r_e \to r_c$, as expected.

## 3. Simulation

We performed two numerical simulations to verify the accuracy of our formula by using the model shown in Fig. 3. The parameters used in the simulation are listed in Table. I. The simulation was performed in axisymmetric 2D to take advantage of cylindrical symmetry of the geometry. First, a sinusoidal electric potential of amplitude $5.0$ mV was applied to the WE without DC bias, while the cytoplasm and the edge face of the model domain ($r = r_c$) were set to zero potential. The frequency was swept from $10^2$ Hz to $10^7$ Hz in logarithmic steps. Second, we calculated impedances from the ratios of the complex representations of the input voltages to those of the simulated currents. In addition, the frequency characteristics of the impedance magnitudes and phases were calculated via the discrete Fourier transform and are shown in a Bode plot. In the following two simulations, $r_c$ was set to $5.0$ μm and an insulation boundary condition was imposed on the substrate. In the first simulation, we swept $r_e$ from $1.0$ μm to $4.0$ μm stepped by $1.0$ μm with $h$ fixed at $100$ nm. In the second simulation, we swept $h$ from $100$ nm to $500$ nm stepped by $50$ nm with $r_e$ fixed at $1.0$ μm. The numerical simulations were performed using COMSOL Multiphysics 5.6, where the differentail forms of Maxwell's equations are solved in this software.

## 4. Results and Discussion

Fig. 4 shows a Bode plot of the simulated impedance at a fixed $h = 100$ nm (dots) and the theoretical impedance calculated using $Z_c$ (solid lines). In Fig. 4, the formula well-describes the simulated impedance over a wide range of frequency, regardless of the value of $r_e$.

Fig. 4 shows that the electrical double-layer capacitance at the interface between the electrode and solution is dominant from $10^2$ Hz to $10^3$ Hz because the phase difference starts with $-90°$. Assuming that the non-dimensional variable $x$ satisfies $x \ll 1$, the modified Bessel functions of the first and the second order are approximated by the following equations

$$I_0(x) \simeq 1,$$
$$I_1(x) \simeq \frac{x}{2}, \tag{43}$$

$$K_0(x) \simeq -\ln\frac{x}{2} - C_E,$$
$$K_1(x) \simeq \frac{1}{x}, \tag{44}$$

where $C_E$ is Euler's constant. At a sufficiently low frequency, at which $\gamma_A r_e \ll 1$ and $\gamma_B r_e \ll 1$, $\Delta$, $C_A$, and $Z_c^{-1}$ can thus be approximated using following equations ($V_c = 0$)

$$\Delta \simeq -\frac{1}{r_e}, \tag{45}$$

$$\begin{aligned}C_A &\simeq -r_e(1-k^2)\big(V_e \gamma_B K_1(\gamma_B r_e)\big) \\ &= -V_e(1-k^2) \\ &= -V_e \frac{Z_m}{Z_n + Z_m},\end{aligned} \tag{46}$$

$$\begin{aligned}Z_c^{-1} &\simeq \frac{2\pi r_e Z_m I_1(\gamma_A r_e)}{Z_n(Z_n + Z_m)\gamma_A} + \pi r_e^2 \frac{1}{Z_n + Z_m} \\ &= \frac{\pi r_e^2}{Z_n + Z_m}\left(\frac{Z_m}{Z_n}\frac{2}{\gamma_A r_e}I_1(\gamma_A r_e) + 1\right) \\ &= \frac{\pi r_e^2}{Z_n + Z_m}\left(\frac{Z_m}{Z_n}\frac{2}{\gamma_A r_e}\frac{\gamma_A r_e}{2} + 1\right) \\ &= \frac{\pi r_e^2}{Z_n}.\end{aligned} \tag{47}$$

Consequently, the impedance of the electrical double-layer capacitance is dominant at sufficiently low frequency, as seen in Fig. 4.

Above $10^6$ Hz, the phase differences become closer to $-90°$ in Fig. 4, thus this frequency range is the region where the capacitance is becoming dominant. This capacitance consists of the electrical double-layer capacitance at the metal/solution interface and the capacitance of the cell membrane. In general, the impedance of the cell membrane ($Z_m$) is much larger than that of the electrical double-layer capacitance ($Z_n$). Therefore, $Z_m$ is dominant at high frequency. Assuming that the non-dimensional variable $x$ satisfies $x \gg 1$, the modified Bessel functions of the first and second orders can be approximated by the following equations

$$I_0(x) \simeq I_1(x) \simeq \sqrt{\frac{1}{2\pi x}} \exp(x), \tag{48}$$

$$K_0(x) \simeq K_1(x) \simeq \sqrt{\frac{\pi}{2x}} \exp(-x). \tag{49}$$

From these approximations, at a sufficiently low frequency at which $\gamma_A r_e \gg 1$ and $\gamma_B r_e \gg 1$, $\Delta$, $C_A$, and $Z_c^{-1}$ can be approximated using following equations ($V_c = 0$)

$$\Delta = \frac{1}{k}\sqrt{\frac{1}{2\pi\gamma_A r_e}}\sqrt{\frac{1}{r_e r_c}} \exp(\gamma_A r_e)[\sinh[\gamma_B(r_e - r_c)] - k \cosh[\gamma_B(r_e - r_c)]], \tag{50}$$

$$\begin{aligned} C_A &\simeq k\sqrt{2\pi\gamma_A r_e} \frac{V_e(1-k^2)\cosh[\gamma_B(r_e - r_c)]}{\exp(\gamma_A r_e)[\sinh[\gamma_B(r_e - r_c)] - k\cosh[\gamma_B(r_e - r_c)]]} \\ &= k\sqrt{2\pi\gamma_A r_e} \frac{V_e(1-k^2)}{\exp(\gamma_A r_e)[\tanh[\gamma_B(r_e - r_c)] - k]} \\ &\simeq k\sqrt{2\pi\gamma_A r_e} V_e \exp(-\gamma_A r_e)\frac{1-k^2}{1-k} \\ &= \sqrt{2\pi\gamma_A r_e}\, V_e k(1+k)\exp(-\gamma_A r_e), \end{aligned} \tag{51}$$

$$\begin{aligned} Z_c^{-1} &\simeq -\frac{2\pi r_e}{Z_n \gamma_A}\sqrt{2\pi\gamma_A r_e}\, k(1+k)\exp(-\gamma_A r_e)\sqrt{\frac{1}{2\pi\gamma_A r_e}}\exp(\gamma_A r_e) + \pi r_e^2 \frac{1}{Z_n + Z_m} \\ &= -\frac{2\pi r_e}{Z_n \gamma_A} k(1+k) + \pi r_e^2 \frac{1}{Z_n + Z_m} \\ &= \pi r_e^2 \frac{1}{Z_n + Z_m}\left[-\frac{2}{r_e \gamma_A}\frac{Z_n + Z_m}{Z_n}k(1+k) + 1\right] \\ &= \pi r_e^2 \frac{1}{Z_n + Z_m}\left[-\frac{2(1+k)}{r_e \gamma_A k} + 1\right] \\ &\simeq \pi r_e^2 \frac{1}{Z_n + Z_m}. \end{aligned} \tag{52}$$

Hence, the sum of $Z_n$ and $Z_m$ is dominant at sufficiently high frequency.

In Fig. 4, at around $10^5$ Hz, the phase difference approaches $-20°$. Thus, the distributed impedance which consists of the solution resistance and $Z_n$ is dominant. Moreover, the slope of $|Z|$ approaches 0 as $r_e$ decreases at about $10^5$ Hz. As the resistance is less dependent on the frequency, this is attributed to an increase in resistance. At around $10^5$ Hz, $Z_m$ is negligible because it is sufficiently large. Therefore, region B consists of only solution resistance. As a consequence, the extension of region B with decrease in $r_e$ increases the solution resistance.

In Fig. 4, errors between the simulated and theoretical impedances are not evident, because Fig. 4 is a log-log graph. Therefore, we evaluated the error rates defined by

$$\text{Error rate [\%]} = \frac{|Z_\text{Simulation}| - |Z_\text{Theory}|}{|Z_\text{Theory}|} \times 100. \tag{53}$$

Fig. 5 shows the frequency characteristics of the error rates calculated using Eq. (53). In Fig. 5, all error rates are less than $1.0\ \%$ below $10^6$ Hz, regardless of the value of $r_e$. Thus, the formula well-describes the cell-substrate impedance. However, at around $10^7$ Hz, the error rate increases. This is attributed to the potential drop on the $z$-axis which the derived formula does not consider. To examine the potential drop on the $z$-axis, the simulated current density on the $z$-axis with respect to the $r$-axis is shown in Fig. 6. This figure shows that the simulated current density on the $z$-axis at high frequency is larger than that at low frequency. This indicates that the potential drop on the $z$-axis at high frequency is larger than that at low frequency. Thus, the error at high frequency in Fig. 5 is attributed to the potential drop on the $z$-axis. Furthermore, the simulated current density on the $z$-axis increased at the edge of electrode ($r = r_e$) due to the edge effect, and hence the increase of current density at the edge in Fig. 6 reflects this effect.

Fig. 7 represents the error rates between the impedances simulated at a fixed $r_e = 1.0\ \mu m$ and the theoretical impedance of Eq. (40). In Fig. 7, the error rates increase as $h$ increases. As with the first simulation, this is attributed to the potential drop on the $z$-axis which the formula does not take into account of. The results in Fig. 5 shows that the error rates between theoretical and simulated impedances are not more than $2.0\,\%$, thus the formula can describe appropriately cell-substrate impedance regardless of the WE radius. The results in Fig. 7 shows that the error caused by the potential drop on the z-axis increases as $h$ increases, thus the formula is appropriate when $h$ is sufficiently small. As the actual cell-to-substrate distance is from $50\,\mathrm{nm}$ to $100\,\mathrm{nm}$ when a cell adheres to an electrode[48,49], the error caused by the potential drop on the $z$-axis might be negligible in actual application of our formula.

## 5. Conclusion

We proposed a formula for describing cell-substrate impedance when the electrode is smaller than a single cell. We performed two simulations to verify the accuracy of the formula. The first simulation revealed that our formula can appropriately describe cell-substrate impedance regardless of the electrode radius. The second simulation showed that when cell-substrate gap is large, the potential drop on the $z$-axis cannot be neglected, and our formula becomes inapplicable. However, for realistic cell-substrate gap, error rates are as small as $2.0\,\%$. Our formula would enable quantitative analysis of cell parameters when the electrode is smaller than the cell.

## Acknowledgment

This work was supported by JSPS KAKENHI Grant Number 19K04539.

## References

1   J. L. Griffin, J. P. Shockcor, Metabolic profiles of cancer cells, Nature Reviews Cancer, 4 (2004) 551, https://doi.org/10.1038/nrc1390.

2   M. Rudin, R. Weissleder, Molecular imaging in drug discovery and development, Nature

Reviews Drug Discovery, 2 (2003) 123, https://doi.org/10.1038/nrd1007.

3   R. Edmondson, J. J. Broglie, A. F. Adcock, L. Yang, Three-dimensional cell culture systems and their applications in drug discovery and cell-based biosensors, ASSAY and Drug Development Technologies, 12 (2014) 207, https://doi.org/10.1089/adt.2014.573.

4   M. Oliveira, P. Conceição, K. Kant, A. Ainla, L. Diéguez, Electrochemical sensing in 3D cell culture models: New Tools for developing better cancer diagnostics and treatments, Cancers, 13 (2021) 1381, https://doi.org/10.3390/cancers13061381.

5   C. C. Moura, R. S. Tare, R. O. Oreffo, S. Mahajan, Raman spectroscopy and coherent anti-stokes raman scattering imaging: Prospective tools for monitoring skeletal cells and skeletal regeneration, Journal of The Royal Society Interface, 13 (2016) 20160182, https://doi.org/10.1098/rsif.2016.0182.

6   S. Aknoun, M. Yonnet, Z. Djabari, F. Graslin, M. Taylor, T. Pourcher, B. Wattellier, P. Pognonec, Quantitative phase microscopy for non-invasive live cell population monitoring, Scientific Reports, 11 (2021) 4409, https://doi.org/10.1038/s41598-021-83537-x.

7   S. Yan, D. Yuan, Continuous microfluidic 3D focusing enabling microflow cytometry for single-cell analysis, Talanta, 221 (2021) 121401, https://doi.org/10.1016/j.talanta.2020.121401.

8   J. Y. Chen, L. S. Penn, J. Xi, Quartz crystal microbalance: Sensing cell-substrate adhesion and beyond, Biosensors and Bioelectronics, 99 (2018) 593, https://doi.org/10.1016/j.bios.2017.08.032.

9   X. Zhang, J. Fang, L. Zou, Y. Zou, L. Lang, F. Gao, N. Hu, P. Wang, A novel sensitive cell-based love wave biosensor for marine toxin detection, Biosensors and Bioelectronics, 77 (2016) 573, https://dx.doi.org/10.1016/j.bios.2015.07.062.

10  S. H. Jeong, D. W. Lee, S. Kim, J. Kim, B. Ku, A study of electrochemical biosensor for analysis of three-dimensional (3D) cell culture, Biosensors and Bioelectronics, 35 (2012) 128, https://doi.org/10.1016/j.bios.2012.02.039.

11  J. Hong, K. Kandasamy, M. Marimuthu, C. S. Choi, S. Kim, Electrical cell-substrate impedance sensing as a non-invasive tool for cancer cell study, The Analyst, 136 (2011) 237, https://doi.org/10.1039/c0an00560f.

12  T. Yoshinobu, M. J. Schöning, Light-addressable potentiometric sensors for cell monitoring and Biosensing, Current Opinion in Electrochemistry, 28 (2021) 100727, https://doi.org/10.1016/j.coelec.2021.100727.

13  Y. Xu, X. Xie, Y. Duan, L. Wang, Z. Cheng, J. Cheng, A review of impedance measurements of whole cells, Biosensors and Bioelectronics, 77 (2016) 824, https://doi.org/10.1016/j.bios.2015.10.027.

14  R. A. Nascimento, R. E. Özel, W. H. Mak, M. Mulato, B. Singaram, N. Pourmand, Single cell "Glucose Nanosensor" Verifies Elevated Glucose Levels in Individual Cancer Cells, Nano Letters, 16 (2016) 1194, https://doi.org/10.1021/acs.nanolett.5b04495.

15  T. Kajisa, Y. Yanagimoto, A. Saito, T. Sakata, Biocompatible Poly(catecholamine)-film electrode for potentiometric cell sensing, ACS Sensors, 3 (2018) 476, https://doi.org/10.1021/acssensors.7b00897.

16  T. Sakata, H. Sugimoto, A. Saito, Live monitoring of Microenvironmental ph based on extracellular acidosis around cancer cells with cell-coupled gate ion-sensitive field-effect transistor, Analytical Chemistry, 90 (2018) 12731, https://doi.org/10.1021/acs.analchem.8b03070.

17  S.-M. Kim, N. Kim, Y. Kim, M.-S. Baik, M. Yoo, D. Kim, W.-J. Lee, D.-H. Kang, S. Kim, K. Lee, M.-H. Yoon, High-performance, polymer-based direct cellular interfaces for electrical stimulation and recording, NPG Asia Materials, 10 (2018) 255, https://doi.org/10.1038/s41427-018-0014-9.

18  X. Du, L. Wu, J. Cheng, S. Huang, Q. Cai, Q. Jin, J. Zhao, Graphene microelectrode arrays for neural activity detection, Journal of Biological Physics, 41 (2015) 339, https://doi.org/10.1007/s10867-015-9382-3.

19  G. Tomagra, F. Picollo, A. Battiato, B. Picconi, S. De Marchis, A. Pasquarelli, P. Olivero, A. Marcantoni, P. Calabresi, E. Carbone, V. Carabelli, Quantal release of dopamine and action potential firing detected in midbrain neurons by multifunctional diamond-based microarrays, Frontiers in Neuroscience, 13 (2019) 1, https://doi.org/10.3389/fnins.2019.00288.

20  M. Lian, X. Chen, Y. Lu, W. Yang, Self-assembled peptide hydrogel as a smart Biointerface for enzyme-based electrochemical biosensing and Cell Monitoring, ACS Applied Materials & Interfaces, 8 (2016) 25036, https://doi.org/10.1021/acsami.6b05409.

21  Y. Shu, Q. Lu, F. Yuan, Q. Tao, D. Jin, H. Yao, Q. Xu, X. Hu, Stretchable electrochemical biosensing platform based on Ni-MOF composite/au nanoparticle-coated carbon nanotubes for real-time monitoring of dopamine released from living cells, ACS Applied Materials & Interfaces, 12 (2020) 49480, https://doi.org/10.1021/acsami.0c16060.

22  J.-X. Liu, X.-L. Liang, F. Chen, S.-N. Ding, Ultrasensitive amperometric cytosensor for drug evaluation with monitoring early cell apoptosis based on CU2O@PTPD nanocomposite as signal amplified label, Sensors and Actuators B: Chemical, 300 (2019) 127046, https://doi.org/10.1016/j.snb.2019.127046.

23  Z. Zhang, T. Zheng, R. Zhu, Long-term and label-free monitoring for osteogenic differentiation of mesenchymal stem cells using force sensor and impedance measurement, Journal of Materials Chemistry B, 8 (2020) 9913, https://doi.org/10.1039/d0tb01968b.

24  H. T. Ngoc Le, J. Kim, J. Park, S. Cho, A review of electrical impedance characterization of cells for label-free and real-time assays, BioChip Journal, 13 (2019) 295, https://doi.org/10.1007/s13206-019-3401-6.

25  C. Tong, B. Shi, X. Xiao, H. Liao, Y. Zheng, G. Shen, D. Tang, X. Liu, An annexin V-based biosensor for quantitatively detecting early apoptotic cells, Biosensors and Bioelectronics, 24 (2009) 1777, https://doi.org/10.1016/j.bios.2008.07.040.

26  P. Kenchetty, S. Uno, Impact of width and spacing of interdigitated electrode on impedance-based living cell monitoring studied by computer simulation, Japanese Journal of Applied

Physics, 59 (2019) SDDE02, https://doi.org/10.7567/1347-4065/ab4eb0.

27  M. Ibrahim, J. Claudel, D. Kourtiche, M. Nadi, Geometric parameters optimization of planar interdigitated electrodes for bioimpedance spectroscopy, Journal of Electrical Bioimpedance, 4 (2013) 13, https://doi.org/10.5617/jeb.304.

28  S. Zhu, X. Zhang, M. Chen, D. Tang, Y. Han, N. Xiang, Z. Ni, An easy-fabricated and disposable polymer-film microfluidic impedance cytometer for cell sensing, Analytica Chimica Acta, 1175 (2021) 338759, https://doi.org/10.1016/j.aca.2021.338759.

29  Y.-C. Chuang, K.-C. Lan, K.-M. Hsieh, L.-S. Jang, M.-K. Chen, Detection of glycated hemoglobin (hba1c) based on impedance measurement with parallel electrodes integrated into a microfluidic device, Sensors and Actuators B: Chemical, 171-172 (2012) 1222, https://doi.org/10.1016/j.snb.2012.06.084.

30  H.-G. Jahnke, A. Mewes, F. D. Zitzmann, S. Schmidt, R. Azendorf, A. A. Robitzki, Electrochemical live monitoring of tumor cell migration out of micro-tumors on an innovative multiwell high-dense microelectrode array, Scientific Reports, 9 (2019) 13875, https://doi.org/10.1038/s41598-019-50326-6.

31  A. T. Young, V. A. Pozdin, M. Daniele, In-line microelectrode arrays for impedance mapping of Microphysiological Systems, 2020 IEEE SENSORS, (2020) 1, https://doi.org/10.1109/SENSORS47125.2020.9278636.

32  J. Chung, Y. Chen, S.-J. Kim, High-density impedance-sensing array on complementary metal-oxide-semiconductor circuitry assisted by negative dielectrophoresis for single-cell-resolution measurement, Sensors and Actuators B: Chemical, 266 (2018) 106, https://doi.org/10.1016/j.snb.2018.03.113.

33  V. Viswam, R. Bounik, A. Shadmani, J. Dragas, C. Urwyler, J. A. Boos, M. E. Obien, J. Muller, Y. Chen, A. Hierlemann, Impedance spectroscopy and electrophysiological imaging of cells with a high-density CMOS microelectrode array system, IEEE Transactions on Biomedical Circuits and Systems, 12 (2018) 1356,

https://doi.org/10.1109/TBCAS.2018.2881044.

34 S. Kumashi, D. Jung, J. Park, S. Tejedor-Sanz, S. Grijalva, A. Wang, S. Li, H. C. Cho, C. Ajo-Franklin, H. Wang, A CMOS multi-modal electrochemical and impedance cellular sensing array for massively paralleled Exoelectrogen screening, IEEE Transactions on Biomedical Circuits and Systems, 15 (2021) 221, https://doi.org/10.1109/TBCAS.2021.3068710.

35 Y. Chen, C. C. Wong, T. S. Pui, R. Nadipalli, R. Weerasekera, J. Chandran, H. Yu, A. R. A. Rahman, CMOS high density electrical impedance biosensor array for tumor cell detection, Sensors and Actuators B: Chemical, 173 (2012) 903, https://doi.org/10.1016/j.snb.2012.07.024.

36 U. Kasiviswanathan, S. Poddar, C. Kumar, S. Jit, S. K. Mahto, N. Sharma, A portable standalone wireless electric cell-substrate impedance sensing (ECIS) system for assessing dynamic behavior of mammalian cells, Journal of Analytical Science and Technology, 11 (2020) 25, https://doi.org/10.1186/s40543-020-00223-9.

37 M. Wei, R. Zhang, F. Zhang, N. Yang, Y. Zhang, G. Li, How to choose a proper theoretical analysis model based on cell adhesion and nonadhesion impedance measurement, ACS Sensors, 6 (2021) 673, https://doi.org/10.1021/acssensors.0c02710.

38 S. Tonello, A. Bianchetti, S. Braga, C. Almici, M. Marini, G. Piovani, M. Guindani, K. Dey, L. Sartore, F. Re, D. Russo, E. Cantù, N. Francesco Lopomo, M. Serpelloni, E. Sardini, Impedance-based monitoring of mesenchymal stromal cell three-dimensional proliferation using aerosol jet printed sensors: A Tissue Engineering Application, Materials, 13 (2020) 2231, https://doi.org/10.3390/ma13102231.

39 Y. Zhou, S. Basu, E. Laue, A. A. Seshia, Single cell studies of Mouse Embryonic Stem Cell (MESC) differentiation by electrical impedance measurements in a microfluidic device, Biosensors and Bioelectronics, 81 (2016) 249, https://doi.org/10.1016/j.bios.2016.02.069.

40 S.-B. Huang, Y. Zhao, D. Chen, S.-L. Liu, Y. Luo, T.-K. Chiu, J. Wang, J. Chen, M.-H. Wu,

Classification of cells with membrane staining and/or fixation based on cellular specific membrane capacitance and cytoplasm conductivity, Micromachines, 6 (2015) 163, https://doi.org/10.3390/mi6020163.

41  F. Hempel, J. K. Law, T. C. Nguyen, R. Lanche, A. Susloparova, X. T. Vu, S. Ingebrandt, PEDOT:PSS organic electrochemical transistors for electrical cell-substrate impedance sensing down to single cells, Biosensors and Bioelectronics, 180 (2021) 113101, https://doi.org/10.1016/j.bios.2021.113101.

42  S. Uno, Electrochemical Impedance Sensor for Non-invasive Living Cell Monitoring toward CMOS Cell Culture Monitoring Platform, 2021 International Symposium on Devices, Circuits and Systems (ISDCS), (2021) 1, https://doi.org/10.1109/ISDCS52006.2021.9397893.

43  M. Shiozawa, S. Uno, Electrochemical Impedance Simulation for Single Cell Analysis Using a Microelectrode, Proceedings of the 14th International Joint Conference on Biomedical Engineering Systems and Technologies, (2021) 114, https://doi.org/10.5220/0010266401140120.

44  I. Giaever, C. R. Keese, Micromotion of mammalian cells measured electrically, Proc. Nail. Acad. Sci. USA, 88 (1991) 7896, https://doi.org/10.1073/pnas.88.17.7896.

45  S. Tanaka, K. Kimura, K. Miyamoto, Y. Yanase, S. Uno, Simulation and experiment for electrode coverage evaluation by electrochemical impedance spectroscopy using parallel facing electrodes, Analytical Sciences, 36 (2020) 853, https://doi.org/10.2116/analsci.19P451.

46  P. P. Kenchetty, T. Miura, S. Uno, Computer simulation for electrochemical impedance of a living cell adhered on the inter-digitated electrode sensors, Japanese Journal of Applied Physics, 58 (2019) SBBG15, https://doi.org/10.7567/1347-4065/ab00f0.

47  I. Ermolina, Y. Polevaya, Y. Feldman, B.-Z. Ginzburg, M. Schlesinger, Study of Normal and Malignant White Blood Cells by Time Domain Dielectric Spectroscopy, IEEE Transactions on Dielectrics and Electrical Insulation, 8 (2001) 253, https://doi.org/10.1109/94.919948.

48  K. Toma, H. Kano, A. Offenhäusser, Label-free measurement of cell–electrode cleft gap distance with high spatial resolution surface plasmon microscopy, ACS Nano, 8 (2014) 12612, https://doi.org/10.1021/nn505521e.

49  S. Dai, T. Yu, J. Zhang, H. Lu, J. Dou, M.Zhang, C. Dong, J. Di, J. Zhao, Real-time and wide-field mapping of cell-substrate adhesion gap and its evolution via surface plasmon resonance holographic microscopy, Biosensors and Bioelectronics, 174 (2021) 112826, https://doi.org/10.1016/j.bios.2020.112826.

**Figure Captions**

**FIG. 1.** Schematic of the Giaever's model. WE is the working electrode, and $r_c$ is the cell radius.

**FIG. 2.** (a) Schematic of single-cell analysis using a microelectrode smaller than the cell. A indicates the area between the cell and electrode, and B indicates the area under the cell except in region A. (b) Relationship between the potential and current over an infinitesimal distance $dr$ of region A in (a). (c) Relationship between the potential and current over an infinitesimal distance $dr$ of region B in (a).

**FIG. 3.** The simulation model in cylindrical coordinate.

**FIG. 4.** Bode plot of the simulated impedance at a fixed $h = 100$ nm (dot plot) and the theoretical impedance (line plot) which is calculated by Eq. (40). Legends indicate WE radius.

**FIG. 5.** Error rate between the simulated impedance and theoretical impedances calculated by Eq. (40), as shown in Fig. 4

**FIG. 6.** (a) Current density on the $z$-axis with respect to the $r$-axis when $r_e = 1.0$ µm. (b) Current density on the $z$-axis with respect to the $r$-axis when $r_e = 4.0$ µm. Legends indicate frequencies.

**FIG. 7.** Error rate between the simulated and theoretical impedances calculated by Eq. (40) at a fixed $r_e = 1.0$ µm. Legends indicate $h$.

**TABLE I.** Parameters used in simulation.

| Parameter | Unit | Value |
|---|---|---|
| Double-layer capacitance per unit area[45] | F/m² | 0.89 |
| Solution relative permittivity[46] |  | 78 |
| Solution conductivity[46] | S/m | 1.5 |
| Cell membrane thickness[45] | nm | 5.0 |
| Cell membrane relative permittivity[45] |  | 5.0 |
| Cell membrane conductivity[47] | S/m | $1.0 \times 10^{-9}$ |

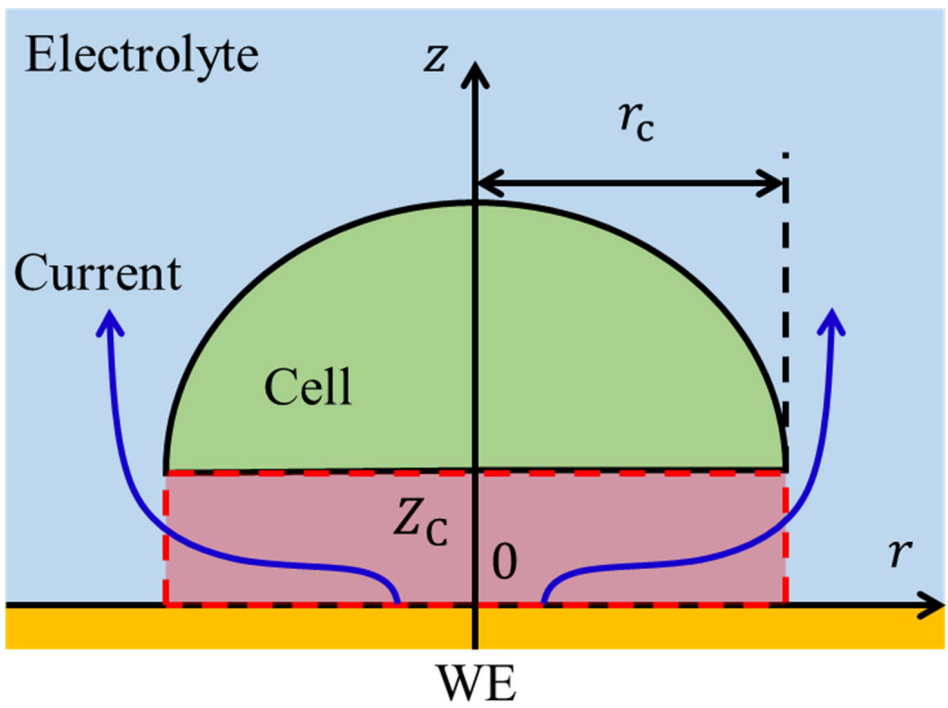

FIG. 1. (Color online)

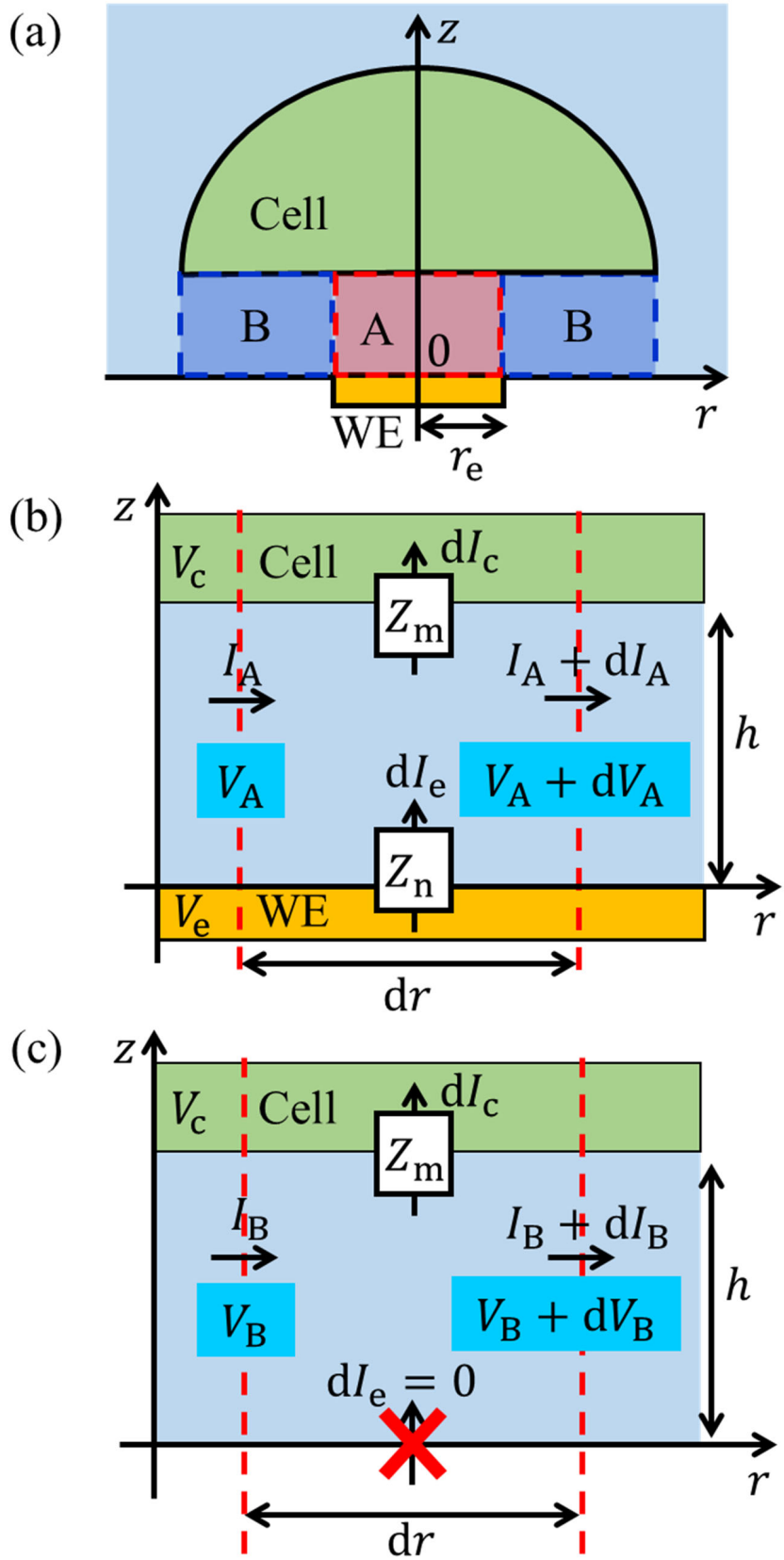

FIG. 2. (Color online)

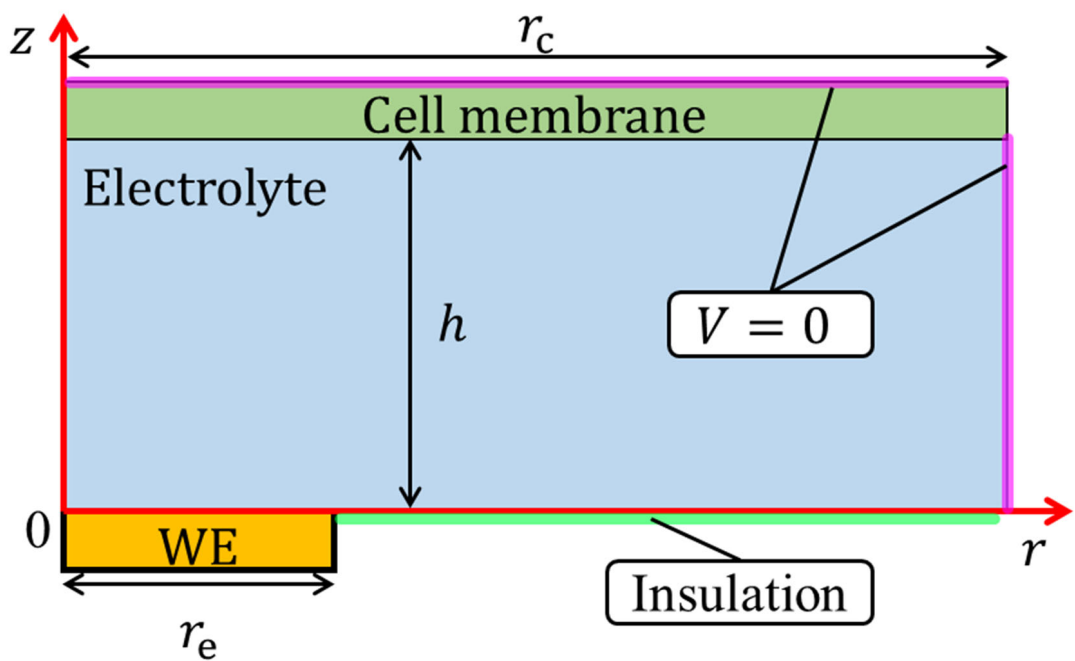

FIG. 3. (Color online)

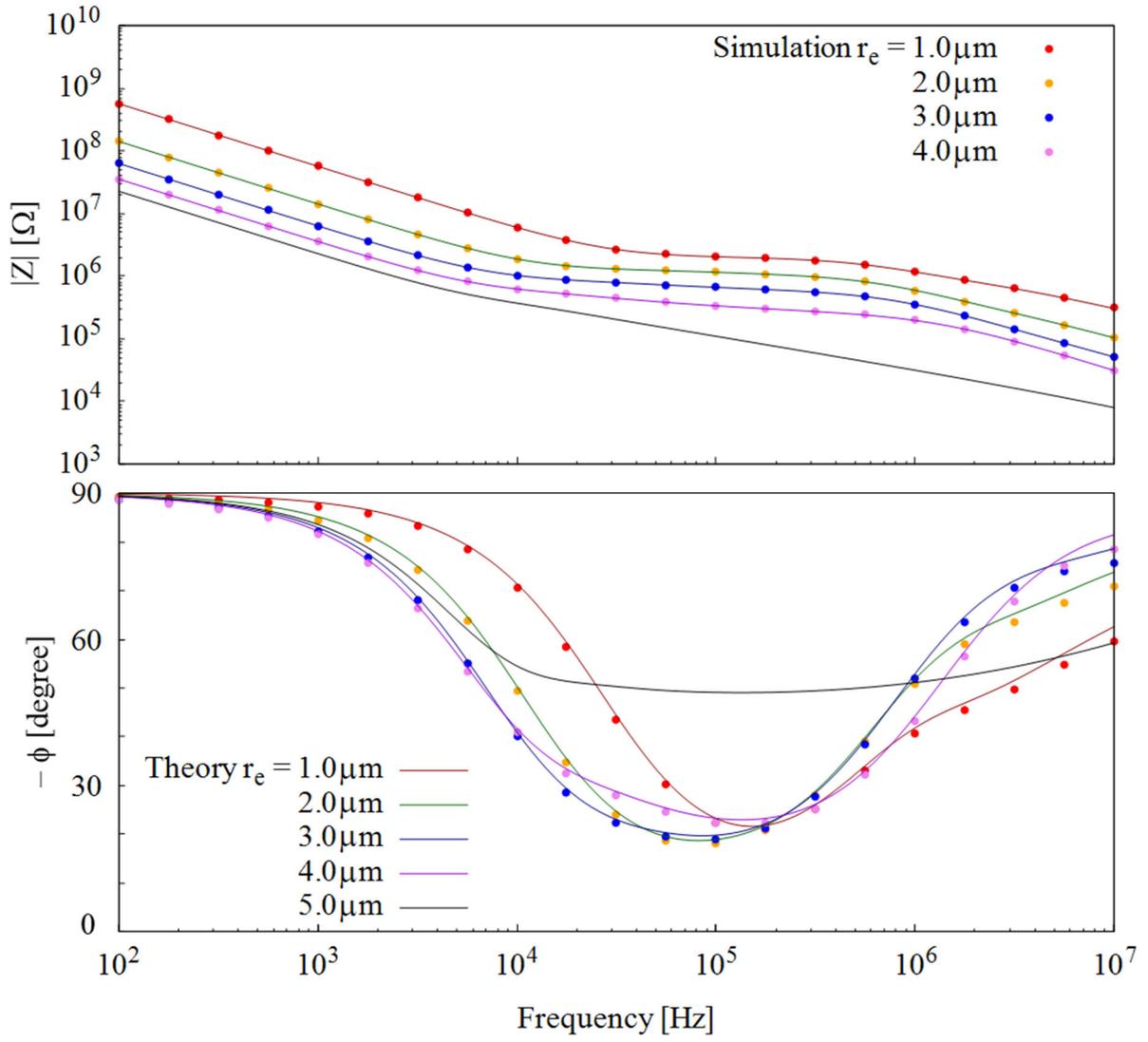

FIG. 4. (Color online)

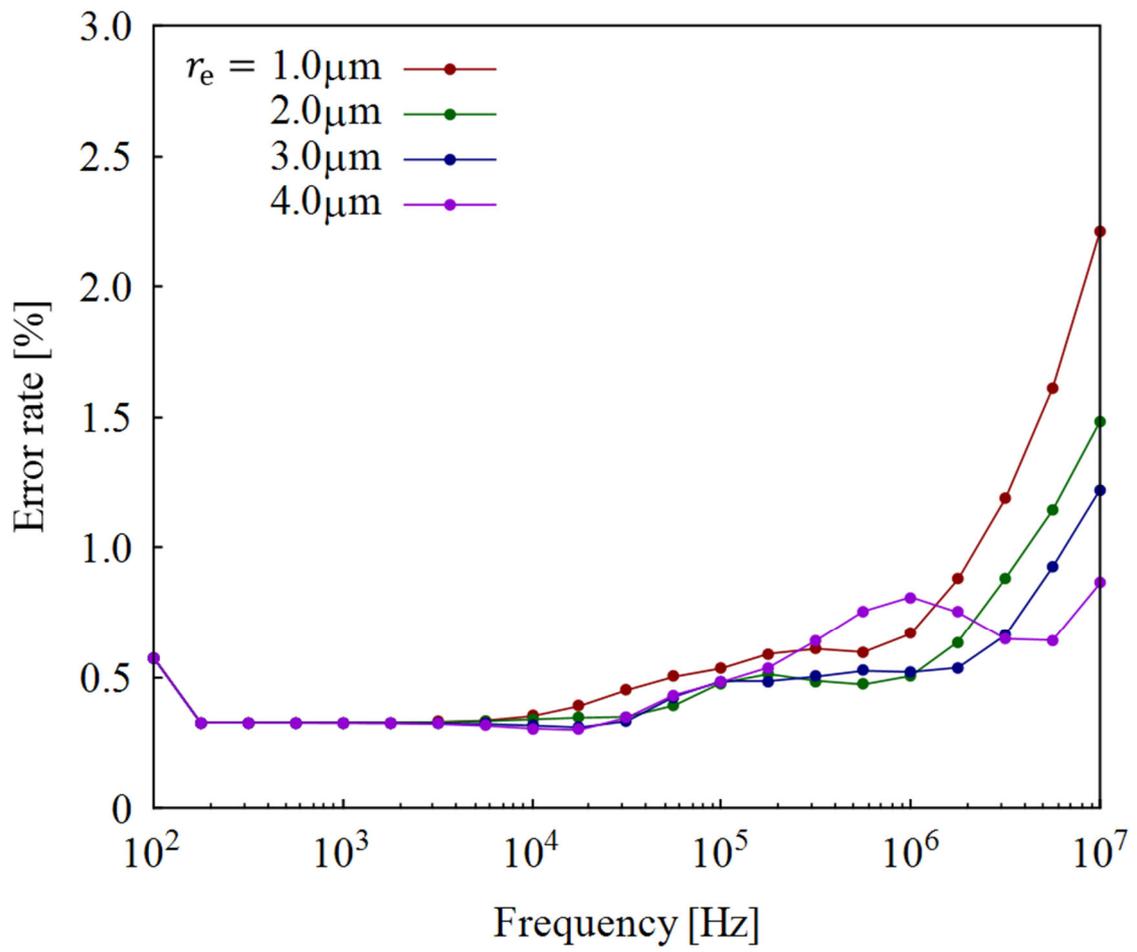

FIG. 5. (Color online)

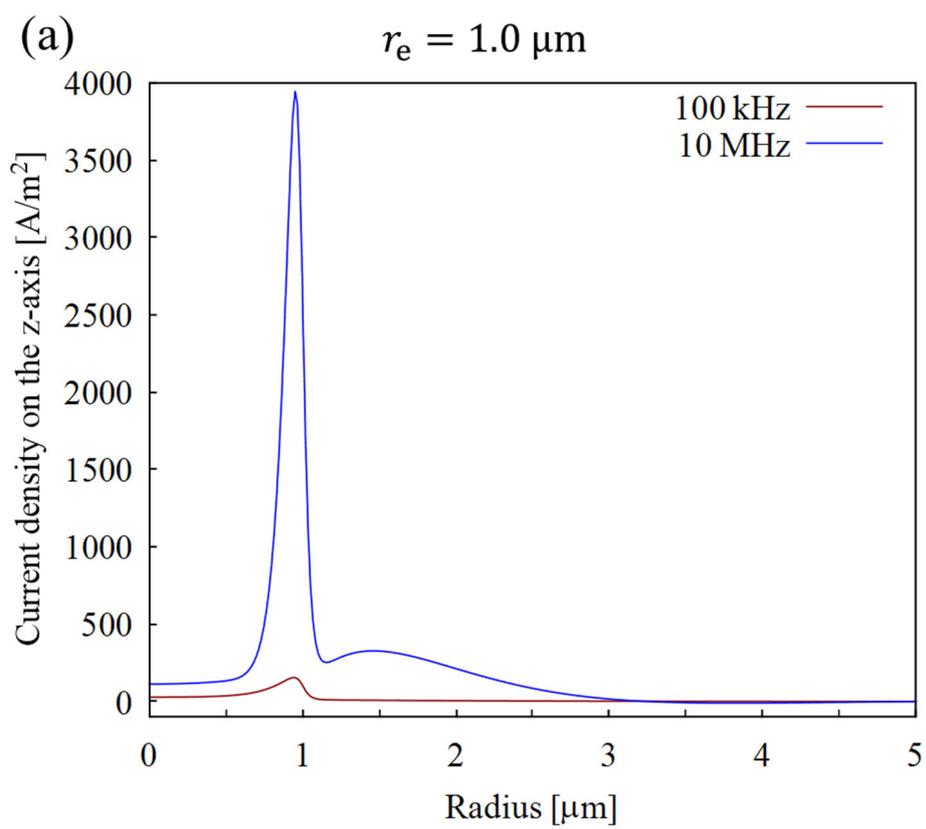

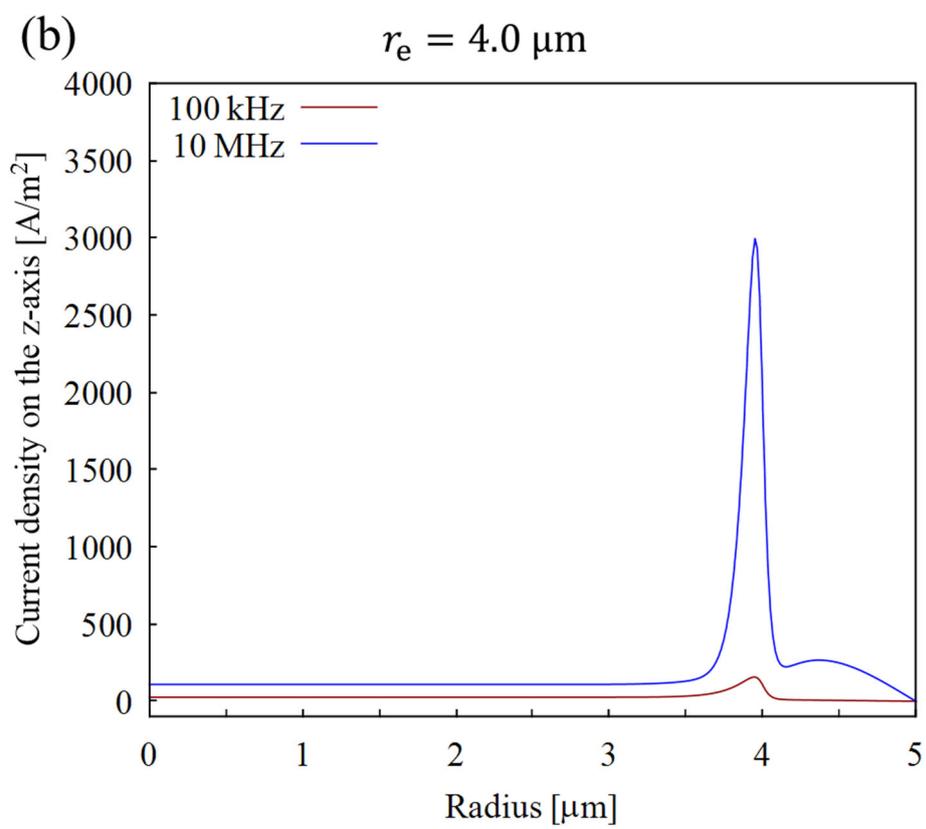

FIG.6. (Color online)

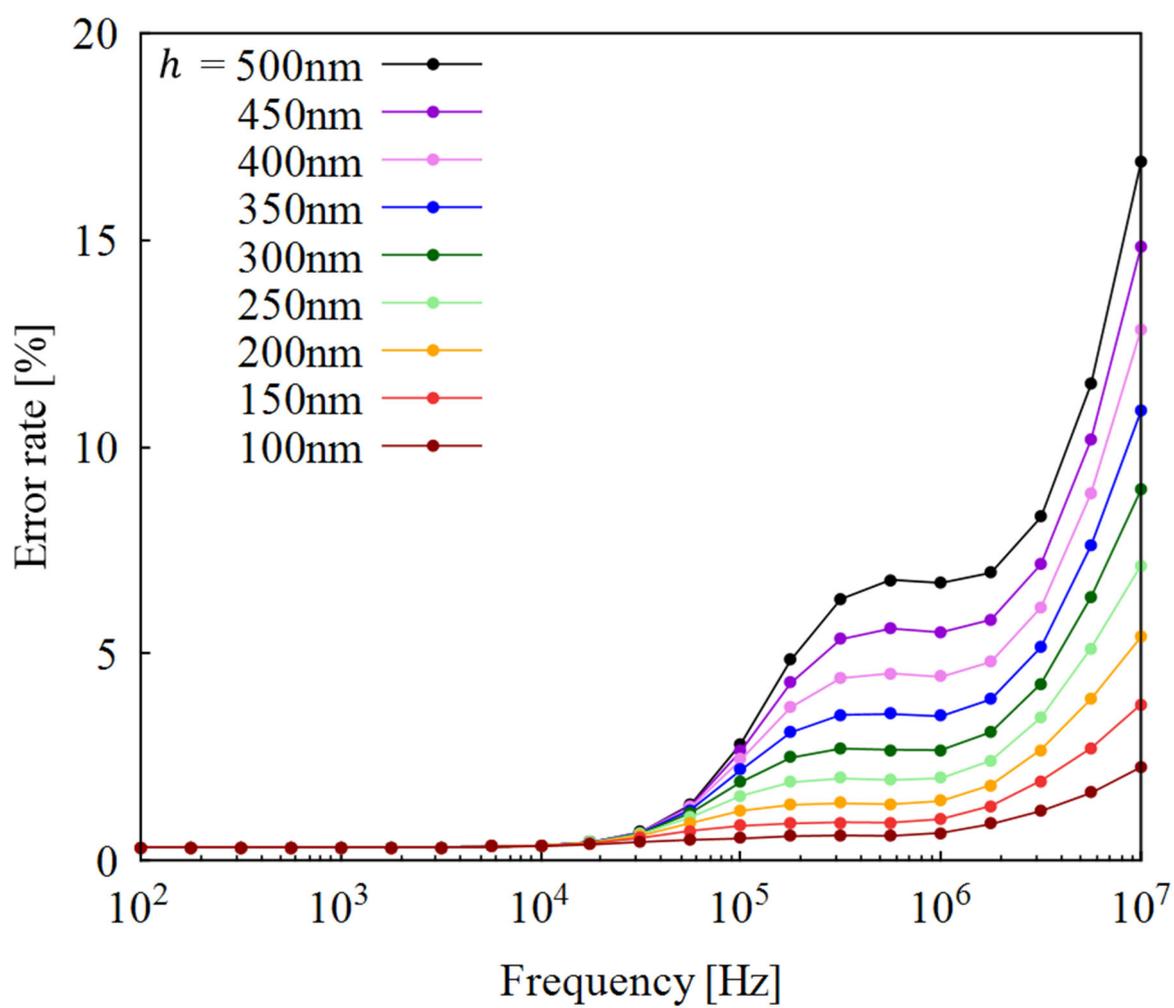

FIG. 7. (Color online)